\documentclass[aps,superscriptaddress,reprint]{revtex4-2}
\usepackage{graphicx}
\usepackage{dcolumn}
\usepackage{bm}
\usepackage{amssymb}
\usepackage{amsmath}
\usepackage{hyperref}
\usepackage{epsf}
\usepackage{xcolor}
\draft
\usepackage{mathtools}
\usepackage{nopageno}
\newcommand{\defeq}{\vcentcolon=}

\begin{document}
\preprint{CDW in 4H_b-TaS2}

\title{Interlayer charge transfer induced by electronic instabilities in the \\ natural van der Waals hetrostructure 4H$_b$-TaS$_2$}
\author{R. Mathew Roy}
\affiliation{1.~Physikalisches Institut, Universit\"at Stuttgart, Pfaffenwaldring 57,
70569 Stuttgart, Germany}
\author{X. Feng}
\affiliation{Max Planck Institute for Chemical Physics of Solids, 01187 Dresden, Germany}
 \author{M. Wenzel}
\affiliation{1.~Physikalisches Institut, Universit\"at Stuttgart, Pfaffenwaldring 57,
70569 Stuttgart, Germany}
  \author{V. Hasse}
\affiliation{Max Planck Institute for Chemical Physics of Solids, 01187 Dresden, Germany}
   \author{C. Shekhar}
\affiliation{Max Planck Institute for Chemical Physics of Solids, 01187 Dresden, Germany}
 \author{M. G. Vergniory}
 \affiliation{Max Planck Institute for Chemical Physics of Solids, 01187 Dresden, Germany}
 \affiliation{Donostia International Physics Center, P. Manuel de Lardizabal 4, 20018 Donostia-San Sebastian, Spain}
 \author{C. Felser}
 \affiliation{Max Planck Institute for Chemical Physics of Solids, 01187 Dresden, Germany}
 \author{A. V Pronin}
 \affiliation{1.~Physikalisches Institut, Universit\"at Stuttgart, Pfaffenwaldring 57,
70569 Stuttgart, Germany}
 \author{M. Dressel}
\affiliation{1.~Physikalisches Institut, Universit\"at Stuttgart, Pfaffenwaldring 57,
70569 Stuttgart, Germany}

\begin{abstract}

The natural van der Waals heterostructure 4H$_b$-TaS$_2$ composed of alternating 1T- and 1H-TaS$_2$ layers serves as a platform for investigating the electronic correlations and layer-dependent properties of novel quantum materials. The temperature evolution of the conductivity spectra $\sigma(\omega)$ obtained through infrared spectroscopy elucidates the influence of band modifications associated with the charge-density-wave (CDW) superlattice on the 1T layer, resulting in a room-temperature energy gap, $\Delta_{\rm CDW}\approx 0.35$~eV.
Supported by density functional theory calculations, we attribute the behavior of interband transitions to the convergence of the layers, which amplifies the charge transfer from the 1T to the 1H layers, progressing as the temperature decreases. This phenomenon leads to an enhanced 
carrier density. The presence of an energy gap and the temperature-tunable charge transfer within the bulk of 4H$_b$-TaS$_2$ --~driven by layer-dependent CDW states~--
contribute to a more comprehensive understanding of other complex compounds of transition-metal dichalcogenides.

\end{abstract}

\date{\today}%
\maketitle


Strongly correlated two-dimensional layers in van der Waals (vdW) materials offer a multitude of electronic and optical applications \cite{ribak2020,novoselov2016,vescoli1998}. In this regard, transition metal dichalcogenides (TMDs) are of great interest because the carrier density can be easily tuned, and the dimension can be reduced by exfoliation \cite{novoselov2016}. 
TMDs also host collective phenomena, including charge-density-waves (CDWs) and superconductivity.   Understanding these quasi-two dimensional (2D) systems may provide insights into other highly-correlated systems such as high-$T_c$ cuprates and twisted bilayer materials \cite{proust2019,devakul2021}.

TMDs are layered materials, in which a transition metal is sandwiched between chalcogen atoms. One popular TMD for exploring CDWs is Ta/Nb$X_2$ ($X$ = S, Se, Te), whose pure polytypes have a trigonal structure of the 1T type or a hexagonal structure of 2H type \cite{dordevic2001,gao2021,hu2022TaTe2}. The spotlight of the present study is on the polytypes of TaS$_2$, which accommodate flat bands and density waves with a chiral nature \cite{guillamon2011,cheng2021, wen2021}. 
However, compounds with alternating layers, such as  4H$_b$-TaS$_2$ or 6R-TaS$_2$, composed of two pure structural types are less explored \cite{wilson1969,meyer1975,lv20236R}. Nevertheless, these natural vdW-heterostructures are excellent platforms for studying interlayer interactions \cite{lv20236R}.

Recent research on 4H$_b$-TaS$_2$ highlights the importance of correlations owing to interlayer coupling and charge transfer in the case of a mesoscopic system; of particular interest are the induced Kondo clusters and flat bands \cite{nayak2023}. Reports of a chiral nature of superconductivity have drawn enormous attention \cite{nayak2023}. These and closely related phenomena are also discussed for another polymorph -- 6R-TaS$_2$ \cite{pal2023, achari2022}.

\begin{figure*}
    \centering
    \includegraphics[width=1.0\textwidth]{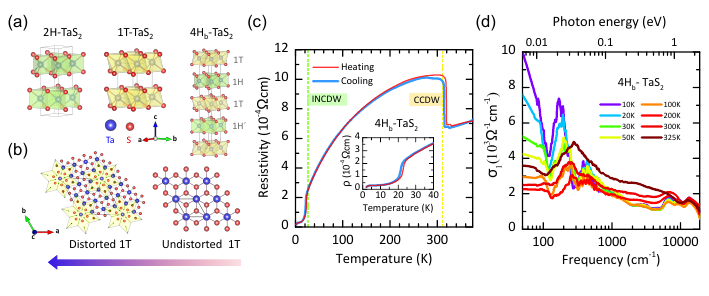}
    \caption{ (a)~Crystal structure of pure and mixed TaS$_2$ polymorphs: 2H-TaS$_2$ with hexagonal structure, 1T-TaS$_2$ with tetragonal structure, and 4H$_b$-TaS$_2$ with alternatively stacked trigonal prismatic 1T and octahedrally coordinated 1H layers of TaS$_2$. (b)~Periodic lattice distortion in the 1T layers forms a star-of David like supercells. (c)~Resistivity of 4H$_b$-TaS$_2$ signals a communsurate CDW (CCDW) and an incommunsurate CDW (INCDW) state at 315 and 24 K respectively, the later is magnified in the insert (note a hysteresis loop). (d)~Real part of the optical conductivity of 4H$_b$-TaS$_2$ for various temperatures; note the two low-energy peaks that sharpen as temperature decreases.}
    \label{resistivity}
\end{figure*}

The crystal structure of 4H$_b$-TaS$_2$, shown in Fig. \ref{resistivity}(a) comprises four layers of alternatively stacked, trigonal prismatic coordinated 1T-TaS$_2$ and octahedrally coordinated 1H-TaS$_2$ (half of 2H) \cite{mattheiss1973}, which are coupled through the vdW force of attraction with an interlayer distance of 5.97~\AA\ \cite{di1973}. 1T-TaS$_2$ in its pure trigonal form is a Mott insulator and experiences a non-commensurate to nearly-commensurate CDW transition at 140 – 180~K \cite{wang2020}. In contrast, 2H-TaS$_2$ in its pure hexagonal coordinated phase exhibits a CDW at around 75~K, and is a metal that becomes superconducting with $T_c$ of at least 0.8~K \cite{wilson1974}. Notably, the layers independently undergo CDW transitions within the bulk \cite{narayan1976}. Unlike 1T-TaS$_2$, 4H$_b$-TaS$_2$ does not exhibit Mott physics, although it retains its 2D characteristics \cite{almoalem2024charge}. The 4H$_b$-TaS$_2$ structure combines the properties of both the 1T and 1H phases, see Fig. \ref{resistivity}, creating a fascinating platform for studying correlations and topological phenomena. All these uniquenesses necessitate the exploration of the intra-and interlayer charge dynamics and its microscopic origin.

The optical properties of materials that undergo CDW transitions have been investigated extensively  \cite{gruner1988}. CDW formation generally involves electron localization accompanied by lattice distortion, which reduces electronic energy and generates an energy gap at the Fermi level ($E_F$).  For example, in kagome compounds with corner-sharing triangular networks,
an energy gap is observed in nonmagnetic systems such as $A$V$_3$Sb$_5$ ($A$ = K, Rb, Cs) \cite{uykur2022,*uykur2021,*wenzel2022rb} and ScV$_6$Sn$_6$ \cite{Hu2023optical,korshunov2023softening}.  In this case, a CDW gap is identified in the optical conductivity by spectral weight transfer from the low-energy to high-energy region. When there are multiple CDWs, new transitions across the gap appear as multiple peaks in the real part of optical conductivity \cite{hu2014multiplecdw}. Clear optical signatures of the CDW energy gaps might be absent, like in FeGe \cite{wenzel2024} or IrTe$_2$ \cite{fang2013IrTe2}. In the case of pure TaS$_2$ polytypes, the optical conductivity indicates a gap at 45~meV in 2H-TaS$_2$ \cite{hu2007} and a partial gap in 1T-TaS$_2$ at higher energies \cite{gasparov2002,dean2011}, making the mixed-layer compounds an excellent case to look at.

This study utilizes optical spectroscopy to investigate the temperature-dependent optical response of the electronic states associated with the formation of two distinct CDW superlattices on the 1T and 1H layers in bulk 4H$_b$-TaS$_2$. Infrared spectroscopy provides high energy resolution, which enables us to examine the bulk electronic responses and band modifications resulting from the CDW superlattice formation. It provides direct insights into transitions across the Fermi level. Our findings reveal an intriguing temperature evolution of the two pronounced low-energy peaks in the optical conductivity. By employing density functional theory (DFT) calculations, we identified their physical origin: the peaks must be attributed to modifications in the electronic bands of the 1T layer and to the subsequent charge transfer from the 1T to 1H layer. 
The CDW distortion in the 1T layer below 315~K, shown in Fig. \ref{resistivity}(b), also induces an energy gap, while no gap is seen under the 1H distortion. 
Our research contributes to the understanding of the impact of band modifications arising from CDW in quasi-2D layers on the bulk properties of the material.


\begin{figure*} [t!]
    \centering
    \includegraphics[width=0.95\textwidth]{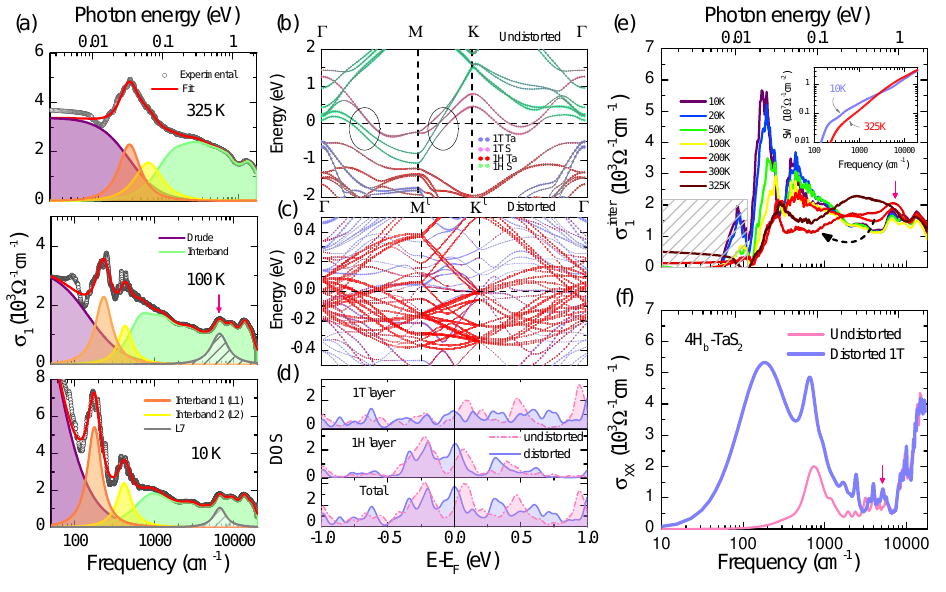}
    \caption{ (a) Decomposed optical conductivity of 4H$_b$-TaS$_2$ at $T = 325$, 100,  and 10~K (top to bottom). The colored shades represent the contributions from the Drude and Lorentz oscillator models, as discussed in the main text and in the Supplimental Material \cite{SM}. (b) Undistorted electronic band structure of the bulk 4H$_b$-TaS$_2$. The regions of relevant interband transitions are highlighted by ellipses. (c) Electronic bands of 4H$_b$-TaS$_2$ with distorted 1T layer. Only the Ta bands are shown. (d) Density of states (DOS) of 1T and 1H layers and the total DOS with and without distortion. (e)~The interband contribution to the optical conductivity. Spectral-weight transfer from high-energy to low-energy peaks occurs, as indicated by the dotted arrows. The inset displays the spectral weight SW as a function of upper bound $\omega_c$. The red arrow indicates the peak that appears at the 315-K CDW transition. (f)~The calculated interband optical conductivity ($\sigma_{xx}$) for the undistorted and distorted cases: the peaks sharpen and move to lower energy under distortion.}
    \label{opticalconductivity}
\end{figure*}

Single crystals of 4H$_b$-TaS$_2$ were grown using a chemical vapor transport method from polycrystalline 4H$_b$-TaS$_2$ powder. The reflectivity from the  $ab$-plane of freshly cleaved crystals with the reflecting plane size approximately $2\times 1~{\rm mm}^2$ and the thickness $\sim$ $0.5 ~{\rm mm}$ was measured in the frequency range 80 to 20\,000~cm$^{-1}$
(i.e.\ 10~meV to 2.5~eV). The complex optical conductivity, $\sigma(\omega) = \sigma_1(\omega) + {\rm i}\sigma_2(\omega)$,  was calculated using the Kramers-Kronig analysis (details in the Supplemental Material \cite{SM}).

The formation of CDW states affects the electrical transport, typically exhibiting a kink or an increase in the dc resistivity $\rho(T)$ owing to the insulating nature of the gapped state.  In Fig. \ref{resistivity}(c), $\rho(T)$ measured within the $ab$-plane demonstrates an abrupt increase at 315~K, associated with the phase transition to a communsurate CDW state in the 1T layer. The temperature-induced periodic lattice distortion in the 1T layer causes the formation of CDW clusters of 13 Ta atoms forming a  $\sqrt{13} \times \sqrt{13}$ star-of-David superstructure with a larger hexagonal unit cell $a = b = 12.04739$~\AA, $c = 23.76717$~\AA\ \cite{nakashizu1986} as illustrated in Fig. \ref{resistivity}(b). In addition, the drop in resistivity at 24~K is associated with an incommensurate CDW in the 1H layer. The hysteresis indicates a first-order transition. The compound also superconducts below $T_c\sim3.7$~K.


Fig. \ref{resistivity}(d) shows the real part of the optical conductivity $\sigma_1(\omega)$ for selected temperatures as derived from the measured reflectivity, plotted in Fig. S1 \cite{SM}. A visible change occurs at around 315~K: a peak appears at 7\,000 cm$^{-1}$ for lower temperatures. Also, two low-energy peaks develop: one at 265~cm$^{-1}$ and another at 450~cm$^{-1}$. These peaks sharpen and blueshift to 180 and 400~cm$^{-1}$, respectively, as the temperature reaches the lowest value (10~K). We note that at 315~K the two peaks are also present but they are broad and strongly overlap with each other. 

\begin{figure*}

    \centering
    \includegraphics[width=1.0\textwidth]{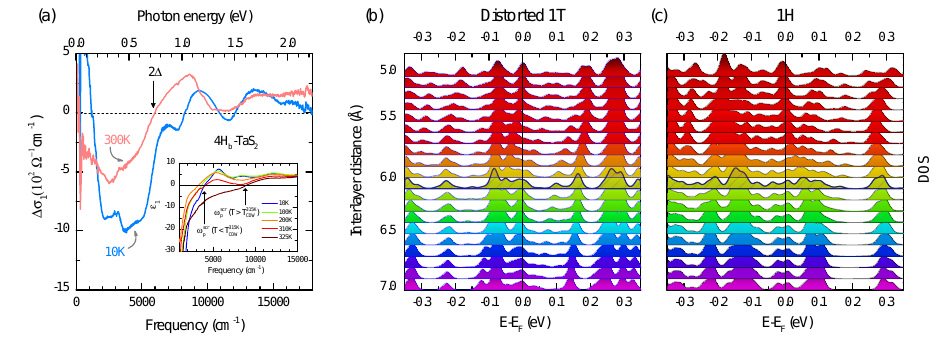}
    \caption{Difference spectra $\Delta\sigma_1$ for 300~K and 10~K (a). Their zero crossing can be taken as an estimate of the energy gap $2\Delta_{\rm{CDW}}$\cite{uykur2021}. The inset in (a) shows the temperature-dependent dielectric permittivity, with the zero crossing being the screened plasma frequency $\omega^{scr}_p$. Calculated interlayer-distance dependence of the DOS for the distorted 1T layer (b) and undistorted 1H layer with distortions present in the 1T layer (c). The interlayer distance was varied from 5 to 7 \AA. There is an enhanced charge transfer from 1T to 1H layer with decreasing layer separation. The bold curve represents the nominal interlayer distance of the bilayer ($\sim$~5.9~\AA).}
    \label{energygapandinterlayer}

\end{figure*}

We decomposed the optical conductivity using the Drude and Lorenz-oscillator models to elucidate the free electron and interband contributions, as described in the Supplemental Material \cite{SM}. The results of this modeling are shown in Fig. \ref{opticalconductivity}(a). The Drude term sharpens with cooling, as the free-carrier scattering rate $1/\tau$ decreases, which, in turn, indicates the persistence of metallicity at low temperatures. The two low-energy interband transitions are shown in orange and yellow. The transitions at higher frequencies were modeled with eight Lorentzians and shown in green. Another interband transition (L7), that forms immediately below 315~K is shaded in grey.

In Fig. \ref{opticalconductivity}(e), the interband portion of optical conductivity [$\sigma^{\rm{inter}}_1(\omega)$] is displayed after subtracting the Drude term from the experimental data. A comparison between 325 and 300~K reveals a spectral-weight transfer to higher energy, indicating the presence of an energy gap (note the L7 peak indicated by the red arrow). Further cooling leads to a dramatic transfer of the spectral weight, ${\rm SW}(\omega_c) = \int_{0}^{\omega_c} \sigma _{1}(\omega){\rm d}\omega$, from relatively high energy ($\sim 4\,000$~cm$^{-1}$) to the two low-energy peaks. This transfer is clearly demonstrated by the calculated SW$(\omega_c)$, as depicted in the inset of Fig. \ref{opticalconductivity}(e).

The energy gap for temperatures below the CDW transition can be determined from the difference spectra, $\Delta \sigma_1 \defeq \sigma_1^{\rm{inter}}(\omega,T<T_{\rm CDW})-\sigma_1^{\rm{inter}}(\omega,T>T_{\rm CDW})$, as illustrated in Fig.~\ref{energygapandinterlayer}(a) (we take the 325-K data as the subtractive). The zero crossing corresponds to the energy gap, $2\Delta_{\rm CDW}$ \cite{uykur2021, kim2023}. For 300~K, $2\Delta_{\rm CDW}$=6\,000~cm$^{-1}$. $\Delta \sigma_1$ becomes negative below this frequency, indicating a significant transfer of SW to high energy owing to the formation of $2\Delta_{\rm CDW}$. This is a result of the 1T layer distortion. At $T=10$~K, the zero-crossing rises to 8\,000~cm$^{-1}$. However, $\Delta \sigma_1$ becomes positive below 1200~cm$^{-1}$, indicating the existence of a low-energy transfer of SW despite a gap. The presence of the two rapidly evolving low-energy transitions [Fig. \ref{opticalconductivity}(a)] suppresses the clear single-particle excitations across the gap with $\sigma_1 \rightarrow 0$. These observations might indicate that the CDW gap exists only at certain momentum directions.

Our estimate of the CDW gap matches well the optical-conductivity results for 1T-TaS$_2$ \cite{barker1975,gasparov2002}, indicating that the observed $\Delta_{\rm{CDW}}$ is mostly related to the CDW formation in the 1T layer of 4H$_b$-TaS$_2$. We obtain $2\Delta/k_{B}T_{\rm{CDW}} \approx 27$, suggesting strong coupling in 4H$_b$-TaS$_2$. We note that the gaps found from STM measurements for both, 1T- \cite{ozaki1983,wang1990} and 4H$_b$-compounds \cite{ozaki1983,tanaka1989}, are systematically below the values reported in optical studies, but the $2\Delta/k_{B}T_{\rm{CDW}}$ ratios remain in the strong-coupling limit.

The inset of Fig.~\ref{energygapandinterlayer}(a) presents the dielectric permittivity $\epsilon_1$, whose 
positive-slope zero crossing yields the screened plasma frequency, $\omega_{p}^{\rm scr}=\omega_p/\sqrt{\epsilon_{\infty}}$; here $\epsilon_{\infty}$ is the high-energy dielectric contribution, which can be approximated as 10 for all temperatures, and $\omega_p$ is the unscreened plasma frequency ($\omega_{p}= \sqrt{4\pi ne^2/m^{*}}$) proportional to the square root of the carrier concentration $n$ and dependent on the effective mass $m^{*}$ \cite{DresselGruner2002}. A significant change in the plasma frequency is observed across $T= 315$~K: 
$\omega^{\rm scr}_p$ decreases from $\sim 8\,800$~cm$^{-1}$ at $T=325$~K to $\sim$ 4\,000 cm$^{-1}$ at 300 K, indicating an almost 60\% reduction in free carriers owing to the CDW formation and the emerging gap in the 1T layer. The free-carrier scattering rate decreases from 800~cm$^{-1}$ at room temperature to approximately 50~cm$^{-1}$ at 10~K, indicating a `more metallic' state at low temperatures. This is consistent with the presence of several ungapped (metallic) bands as discussed in Ref. \cite{vescoli1998}. Hence, the charge excitations are minimally affected by the CDW transition at 24~K, ruling out the possibility of a noticeable gap from the 1H layer, consistent with the reported Raman spectra \cite{nakashizu1984}. However, the strength of the L1 peak at 180~cm$^{-1}$ [Fig. \ref{opticalconductivity}(e)] enhances below $T=25$~K~(see Fig.~S1(c) \cite{SM}). This can be explained by formation of a small CDW gap at below 100 cm$^{-1}$, which is unresolvable in our experiments. 

In order to comprehend the origin of the evolving low-energy peaks, we employed DFT calculations to obtain band-resolved electronic dispersions in various cases. The calculations were carried out using the Vienna {\it ab initio} Simulation Package (VASP) by including spin orbit coupling. Firstly, we examined the band structure of undistorted bulk  4H$_b$-TaS$_2$ and found that the majority of the 1T contribution is located above the 1H bands, see Fig. \ref{opticalconductivity}(b). Between the parallel bands around the M point (highlighted by ellipses in Fig. ~\ref{opticalconductivity}(b)), transitions across $E_{F}$ should occur at around 0.25 -- 0.75 eV (i.e.\ 2\,000 to 7\,000~cm$^{-1}$). Secondly, when the 1T layer of the bulk 4H$_b$-TaS$_2$ unit cell is subject to distortion, the bands from the 1T layer rise towards $E_{F}$, as depicted in Fig.~\ref{opticalconductivity}(c). In the vicinity of $E_{F}$, this alteration decreases the density of states (DOS) of the 1T bands and increases the DOS of the 1H bands, as illustrated in Fig.~\ref{opticalconductivity}(d). Inclusion of the distortions leads to a redistribution of the total DOS near 0.5 eV: it gets depleted at around this energy, but enhanced at somewhat higher and lower energies, see the bottom frame of Fig. \ref{opticalconductivity}(d).

The calculated optical spectra, plotted Fig.~\ref{opticalconductivity}(f), reveal that the conductivity peaks shift to lower energies and are enhanced when distortions are included in the 1T layer. These results are in accordance with our experimental observations. The temperature evolution of the conductivity spectra can be explained by the enhanced proximity of the layers with decreasing temperature, which results in a charge transfer between the layers. Fermi-surface calculations of Gao {\it et al.}, suggested that charge is transferred out of the 1T layer, since the Fermi surface of the 1T-layer bands, compared to pure 1T-TaS$_2$, is reduced in size in 4H$_b$-TaS$_2$ to accommodate the 4H$_b$ unit cell \cite{gao2020}. A recent ARPES study  reports similar observations \cite{almoalem2024charge}.

A deeper understanding of interlayer interactions and charge transfer is gained by calculating the DOS for a varying interlayer distance between 1T and 1H layers. For convenience, we reduced the unit cell to a 1T/1H-TaS$_2$ bilayer system. A $\sqrt{13} \times \sqrt{13}$ supercell is created for the distorted 1T layer. We restrict the effects of the 1T layer as 1H distortion and interlayer coupling have minimal effects on this bilayer system \cite{almoalem2024charge}. 
By following the DOS near $E_F$ in Fig.~\ref{energygapandinterlayer}(b,c), when the interlayer distance is artificially decreased from its nominal value of $\sim$~5.9~\AA (the bold dark-blue curve) to 5~\AA, we see that the DOS in the distorted 1T layers is reduced, while it is enhanced in the distorted 1H layer.
The opposite trend was observed when the interlayer separation is increased to 7~\AA. Therefore, the overall DOS in the 1H+1T system also increased, when the system is distorted in such a way that the layers are brought closer together (see Fig.~S4 \cite{SM}). This observation, in comparison with our experimental results, demonstrates the high sensitivity of the interlayer interaction. Charge transfer can be enhanced under external pressure \cite{friend1977}. In 6R-TaS$_2$, Wang {\it et al.} observed a 1.12~\AA ~decrease in the $c$-axis at 2~GPa \cite{wang2024}. The highly sensitive interlayer interaction also suppresses the CDW temperature and enhances the $T_c$ \cite{yan2023,wang2024}. 
We also note that localized electronic states can be obtained for the 1T/1H-TaS$_2$ bilayer system, although the nature of these states is debated \cite{vavno2021artificial,crippa2024heavy}.

Our findings from the infrared optical conductivity and DFT analysis indicate that the distortion in the quasi-2D 1T layer of 4H$_b$-TaS$_2$ with two independent CDW states significantly influences the bulk electrodynamic response. 
The CDW distortion in the 1T  layer goes hand in hand with band modifications and with a significant charge transfer from the 1T to the 1H layers. This charge transfer is enhanced when the interlayer distance is reduced, and progresses as the system is cooled down. The heavy-charge background and ungapped metallic bands masked the CDW gap associated with the CDW state from the 1H layer. 
An energy gap of 0.35 eV is detected in the optical conductivity and can be associated with the CDW state in the 1T layer. A pressure-dependent optical study may yield a further insight into the interlayer effects with a larger charge transfer.

The authors acknowledge the technical support from Gabriele Untereiner and discussions with Sudip Pal and Ji Eun Lee. 
The project was supported by the Deutsche Forschungsgemeinschaft (DFG) via DR228/51-3.


\bibliography{4Hb_main}

\newpage

\newpage
\vspace*{-2.0cm}
\hspace*{-2.5cm} {
  \centering
  \includegraphics[width=1.2\textwidth, page=1]{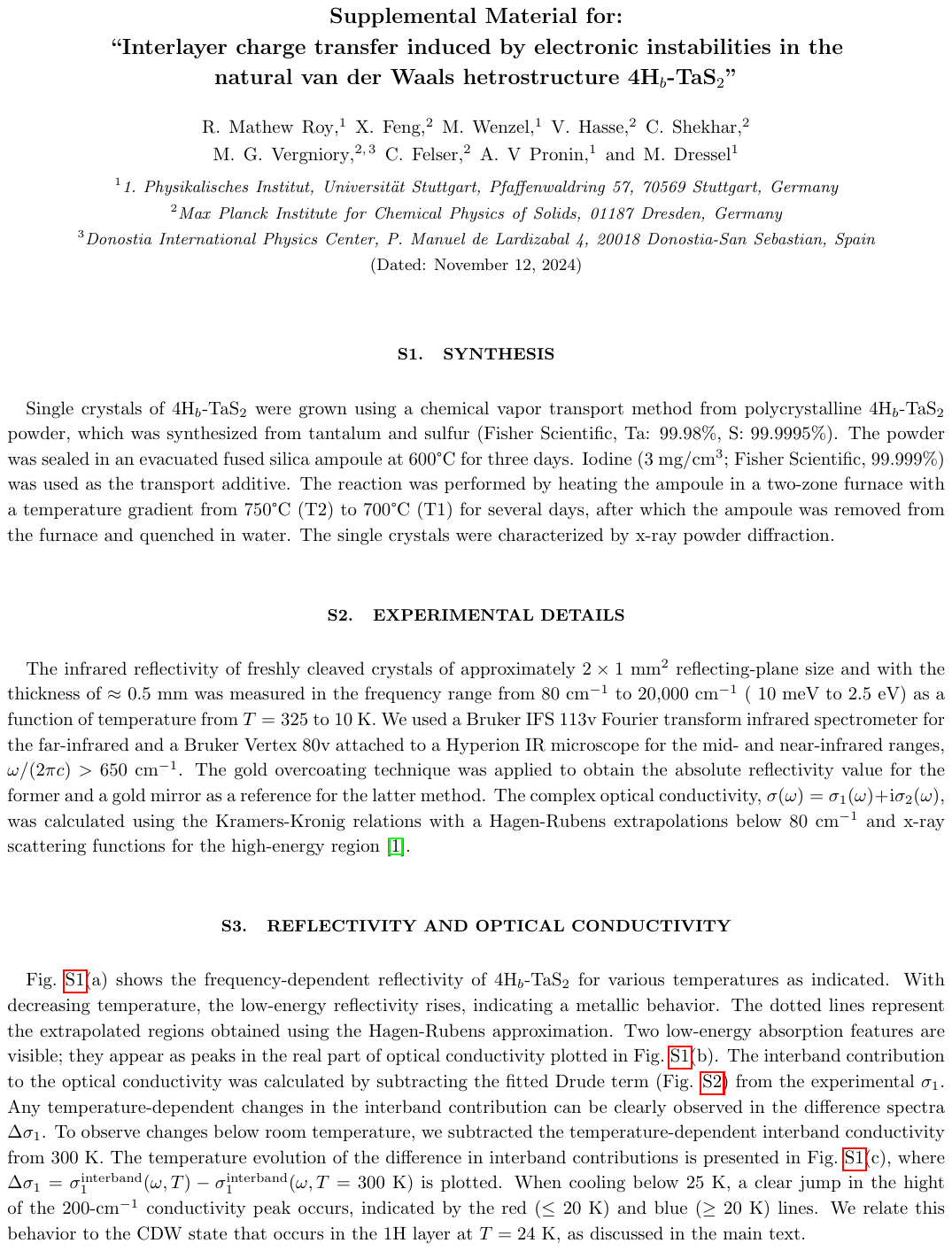} \\
  \ \\
}

\newpage
\vspace*{-2.0cm}
\hspace*{-2.5cm} {
  \centering
  \includegraphics[width=1.2\textwidth, page=2]{sm.pdf} \\
  \ \\
}

\newpage
\vspace*{-2.0cm}
\hspace*{-2.5cm} {
  \centering
  \includegraphics[width=1.2\textwidth, page=3]{sm.pdf} \\
  \ \\
}

\newpage
\vspace*{-2.0cm}
\hspace*{-2.5cm} {
  \centering
  \includegraphics[width=1.2\textwidth, page=4]{sm.pdf} \\
  \ \\
}

\newpage
\vspace*{-2.0cm}
\hspace*{-2.5cm} {
  \centering
  \includegraphics[width=1.2\textwidth, page=5]{sm.pdf} \\
  \ \\
}
       
\end{document}